\documentclass[twocolumn,amssymb, amsmath,nobibnotes,nofootinbib,showpacs, aps, prb]{revtex4-1}
\usepackage[dvips]{graphicx,color}
\begin{document}
%\title{Spin correlated dipole glassiness in CuCrS$_2$: A type {\rm II} multiferroic compound}
\title{Spin correlated dielectric memory and rejuvenation in relaxor ferroelectric CuCrS$_2$}
\author{A. Karmakar$^1$}
\author{K. Dey$^1$}
\author{S. Chatterjee$^2$}
\author{S. Majumdar$^1$}
\author{S. Giri$^{1}$}
%\email{sspsg2@iacs.res.in} 
\affiliation{$^1$Department of Solid State Physics, Indian Association for the Cultivation of Science, Jadavpur, Kolkata 700 032, India \\$^2$UGC-DAE Consortium for Scientific Research, Kolkata Centre, Salt Lake, Kolkata 700 098, India}

\begin{abstract}
CuCrS$_2$, a Heisenberg antiferromagnet with layered edge sharing triangular lattice, exhibits a spiral magnetic order. Dielectric ($\epsilon$) and polarization studies show magnetoelectric (ME) coupling at N\'{e}el temperature ($T_N$=38 K) where simultaneous dielectric and magnetic long range order occur. The compound shows a diffused ferroelectric (FE) transition and slow relaxation dynamics above $T_N$, indicative of relaxor FE behavior. Interestingly, memory effect and magnetic field induced rejuvenation are also observed in $\epsilon$, establishing cooperative glassy dynamics and ME coupling even above $T_N$. We discuss the role of geometrical frustration and metal ligand hybridization for these unusual properties. 
\end{abstract}


\maketitle

%\section{Introduction}

Contemporary research in multiferroics has evidenced tremendous impetus in material science wherein ferroelectric (FE) order occurs concomitantly with magnetic order $-$ the so called spin driven dielectrics or type II multiferroics.\cite{Khomskii} Rapid progress along this direction has unveiled numerous paradigms\cite{Kimura,Hur,Yama,Tani,Seki,Dey,Singh} belonging to this class. Ferroelectricity in most of these materials has its origin in the helimagnetic order that inherits magnetoelectric (ME) coupling $-$ the mutual control of magnetic and dielectric properties. These days, the study of ME coupling is not only confined to the sphere of long-range order but also extended beyond, to the disordered systems, that exhibit glassy dynamics.\cite{Multi, Hemberger} Examples of such dynamical behavior in ME coupled systems are rare and bear fundamental importance, because the underlying mechanism is still inadequately explained, specially in the dielectric sphere. Relaxor ferroics belonging to this class are technologically promising because of enormous electromechanical response and easy polarizability by external magnetic field. A lone example in chalcogenides is the spinel CdCr$_2$S$_4$ that show interesting relaxor FE behavior.\cite{Hemberger} Here, we report another paradigm of relaxor ferroic material in chalcogenides, that exhibits a rare consequence of {\it dielectric memory} and {\it magnetic field induced rejuvenation}.  

CuCrS$_2$ is a Heisenberg antiferromagnet with a strong magnetoelastic effect at $T_N = 38$  K where a structural transition occurs from rhombohedral to monoclinic symmetry.\cite{Rasch} Cr$^{3+}$ ions form layered edge sharing triangular lattice, that give rise to strong geometrical frustration and exhibit spiral magnetic order with an incommensurate magnetic propagation vector.\cite{Rasch, Str-ref, Winten} Being a possible candidate of spin driven ferroelectrics, in analogy to its oxide counterpart, the delafossite CuCrO$_2$,\cite{Seki} having similar layered structure, crystalline anisotropy, and helimagnetic order, this compound is promising for possible multiferroicity. Particularly, the dichalcogenides, being relatively less studied, exhibit acentric $R3m$ space group as compared to centrosymmetric $R\overline{3}m$ space group of dioxides, the former one being more compatible with ferroelectricity.\cite{Singh} Besides, it may be noted that measurement of the dielectric property of CuCrS$_2$ poses an experimental challenge owing to its relatively high conductivity in the family of layered triangular lattice chalcogenides.

In this Letter, we report simultaneous occurrence of long-range antiferromagnetic (AFM) and FE order at the structural transition, signifying ME coupling in CuCrS$_2$. Additionally, we report a broad dielectric transition at higher temperature above $T_N$. The broad temperature response and frequency dependence of the electric order indicate slow dynamics in agreement with relaxor FE behavior. Interestingly, a memory effect and magnetic field induced rejuvenation of the dielectric response are observed above $T_N$. An interesting scenario of unusual {\it cluster} dipole glass or {\it super dipole glass} state is proposed, which is, in fact, responsible for the observed memory and rejuvenation in the dielectric response. We argue that the role of geometrical frustration and metal ligand hybridization appears crucial for the relaxor multiferroicity in CuCrS$_2$.

%\begin{figure}[b]
%\vskip 0.0 cm
%\centering
%\includegraphics[width = 0.8\columnwidth]{Fig1.eps}
%\caption {(Color online) (a) Unit cell of CuCrS$_2$ showing Cu-tetrahedra and Cr-octahedra. (b) shows the layered edge sharing triangular lattice formed by Cr$^{3+}$ and Cu$^+$ ions.} 
%\label{Str}
%\end{figure}
%

%
%\section{Experiment}

Polycrystalline CuCrS$_2$ is synthesized using standard solid state reaction. Powder X-ray diffraction (XRD) data is recorded in a BRUKER axs diffractometer (8D - ADVANCE) using Cu-K$_\alpha$ radiation. X-ray photoemission spectroscopy (XPS) is recorded in a spectrometer of Omicron Nanotechnology.\cite{Karmakar-XPS} Temperature dependent resistivity [$\rho(T)$] and heat capacity [$C_P(T)$] are measured using home built setups. Magnetic susceptibility is measured in SQUID magnetometer (Quantum Design). Dielectric property is studied using LCR meter (Agilent E4980A). Polarization ($P$) is calculated by integrating pyroelectric current recorded in 6517B electrometer (Keithley). For polarization measurement, silver paint is applied on very thin mica sheets attached to the opposite faces of the specimens. This allows us to apply high voltage on the otherwise semiconducting sample and block the background current due to other factors. Magnetodielectric response is measured using a commercial cryogen-free superconducting magnet system (Cryogenic Ltd., UK). 
%The sample is cooled at a particular poling voltage and the leads are shorted at the lowest temperature for 30 mins before measuring pyroelectric current. The measurement is performed in the warming cycle at a ramp rate of 4 K/min on very small and thin specimens (area $\sim$ 0.18 cm$^2$ and thickness $\sim$ 0.1 cm) in the presence of He exchange gas to ensure rapid thermal response and practically no temperature gradient in the sample.

\par
\begin{figure}[t]
\vskip 0.0 cm
\centering
\includegraphics[width = \columnwidth]{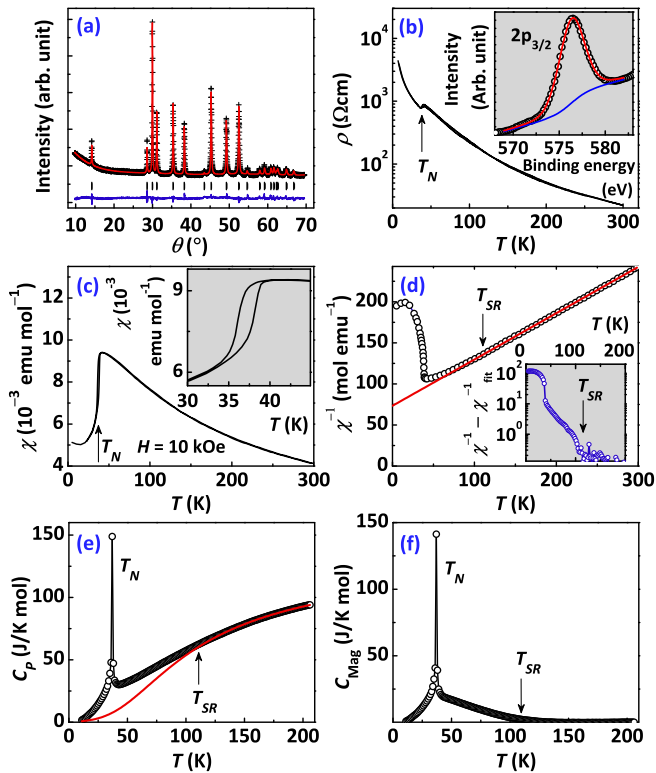}
\caption {(Color online) (a) XRD pattern with Rietveld refinement at room temperature for CuCrS$_2$. (b) $\rho(T)$ plotted in semilog scale. Inset shows XPS spectrum. (c) $\chi(T)$ measured at $H=10$ kOe with the signature of thermal hysteresis (inset) around $T_N$ in enlarged scale. (d) $\chi^{-1}(T)$ displaying fit with Curie-Weiss law. Inset displays the difference plot, $\chi^{-1} - \chi^{-1}_{\rm fit}$ in semilog scale highlighting $T_{SR}$. (e) $C_P(T)$ with line curve displaying fit using Debye model. (f) Magnetic contribution to the specific heat capacity.} 
\label{Char}
\end{figure}

Rietveld refinement of the room temperature XRD pattern ensures phase purity of the sample with no detectable secondary phase, as shown in Fig. \ref{Char}(a). The structure is composed of layered arrangements of Cu$^+$ cations intercalated between two Cr$^{3+}$ layers which form edge sharing triangular lattice. The nearest neighbour interlayer Cr$-$Cr distance of 6.558 {\AA} is larger than the intralayer distance of 3.485 {\AA} corroborating 2D-type Cr-layers. The XPS spectrograph of Cr(2p$_{3/2}$) contribution is shown in the inset of Fig. \ref{Char}(b). A good fit of the Cr(2p$_{3/2}$) peak at 576.35 eV is obtained with a single peak of 80\% Gaussian and 20\% Lorentzian contributions, as displayed by the continuous curve. The curve at the base is the Shirley background. The peak value is close to the values reported for Cr$^{3+}$ states.\cite{XPS-ref} XPS study thus confirms Cr$^{3+}$ state. In Fig. \ref{Char}(b), $\rho(T)$ shows a monotonic increase except for a peak at $\sim$ 40 K followed by a dip at 37.5 K coresponding to the structural transition at $T_N$.\cite{Rasch, Tewari} The data shows a small thermal hysteresis around $T_N$ (not shown for brevity) emphasizing the first order nature of the transition. 

\par
The $T$-dependent dc magnetic susceptibility ($\chi$) measured at $H = 10$ kOe is shown in Fig. \ref{Char}(c). Transition from an apparently paramagnetic (PM) to AFM state occurs at the inflection point ($T_N = 38$ K) in the zero-field cooled (ZFC) curve. Thermal hysteresis observed in the field-cooled (FC) susceptibility is enlarged in the inset, further emphasizing the first order transition at $T_N$. Inverse magnetic susceptibility [$\chi^{-1}(T)$] is plotted in Fig. \ref{Char}(d). The continuous curve, indicating the fit ($\chi^{-1}_{\rm fit}$) of the high-$T$ linear part to Curie-Weiss law, provides $\Theta_{CW} = -130$ K. The effective PM moment obtained from the fit is $\mu_{eff} = 3.77 \mu_B$/f.u. which is close to the spin only value $\mu_{S=3/2} = 3.87 \mu_B$/f.u. The ratio, $\left|\Theta_{CW}\right|/T_N$, is $\sim$ 3.42 indicating strong magnetic frustration.\cite{Rasch} The inset shows the difference between the calculated and the experimental $\chi^{-1}(T)$. The difference plot shows non-zero value at $\sim$ 110 K ($T_{SR}$), below which it rises considerably along with a sharp increase at $T_N$ as $T$ decreases. Figure \ref{Char}(e) shows the specific heat capacity ($C_P$) as a function of $T$. A sharp peak is observed at 37.5 K which marks the structural transition from a high-$T$ rhombohedral to low-$T$ monoclinic phase.\cite{Rasch, Tewari2} The magnetic contribution to $C_P$ ($C_{Mag}$), shown in Fig. \ref{Char}(f), is calculated by subtracting the lattice contribution ($C_{Lat}$) from $C_P(T)$. $C_{Lat}$, depicted by the continuous curve in Fig. \ref{Char}(d), is calculated by fitting the high-$T$ $C_P(T)$ data to the Debye model using least square fit that provides Debye temperature, $\Theta_D = 416$ K. $C_{Mag}(T)$ shows non-zero value below $T_{SR}$ ($\sim$ 110 K) and rises steadily until $T_N$ below which a sharp jump occurs. 

\par
\begin{figure}[t]
\vskip 0.0 cm
\centering
\includegraphics[width = 0.95\columnwidth]{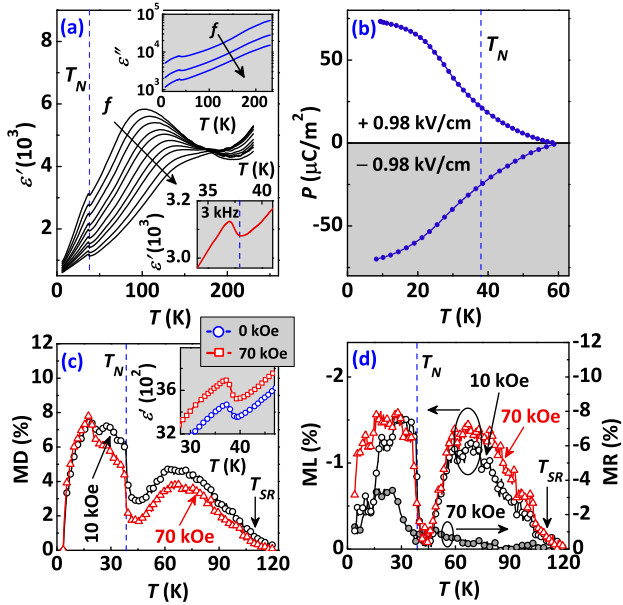}
\caption {(Color online) (a) $\epsilon^\prime(T)$ for $f = 4 -$400 kHz (increasing along arrow direction). Upper inset shows $\epsilon^{\prime\prime}(T)$ for $f = 0.4-$2 MHz. Lower inset shows an enlarged peak at $T_N$. Thermal dependence of (b) $P(T)$ and (c) magnetodielectric effect (MD\%). (d) Left axis: Magnetoloss (ML\%) and (d) Right axis: Magnetoresistance (MR\%). Inset of (c) shows $\epsilon^\prime(T)$ close to $T_N$ at $H$ = 0 and 70 kOe.}
\label{Dielec}
\end{figure}
\begin{figure}[t]
\vskip 0.0 cm
\centering
\includegraphics[width = \columnwidth]{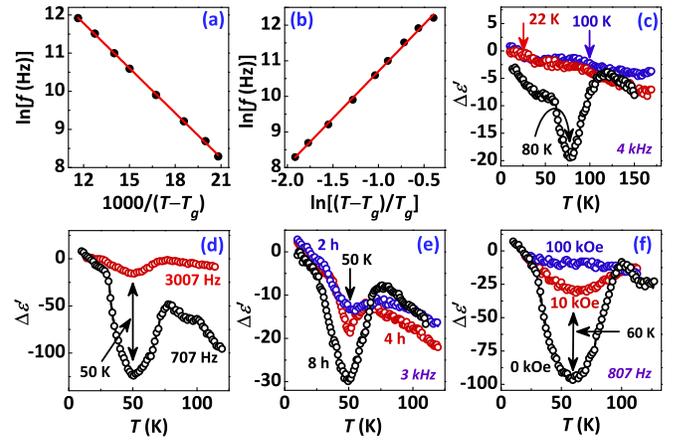}
\caption {(Color online) The $f$-dependence of $T_m$ plotted according to VF (a) and DS (b) laws. $\Delta\epsilon^\prime(T)$ at different (c), (d), (e), and (f) $T_w$ (c), $f$ (d), $t_w$ (e), and H (f).} 
\label{Mem}
\end{figure}

Thermal variations of real ($\epsilon^\prime$) and imaginary ($\epsilon^{\prime\prime}$) components of permittivity are displayed in Fig. \ref{Dielec}(a). $\epsilon^\prime(T)$ shows steep increase with temperature and is superposed by a peak at $\sim$ 37 K, close to $T_N$. This peak ($T_{FE}$) is highlighted in the lower inset for frequency, $f = 3$ kHz, and is characterized by no $f$-dispersion. Such a peak in $\epsilon^\prime(T)$ indicates long-range order of electric dipole moments. The  coincidence of $T_{FE}$ and $T_N$ indicates ME coupling and designates the material as a type II multiferroic material.\cite{Khomskii} Signature of $T_{FE}$ is also indicated by a hump in $\epsilon^{\prime\prime}(T)$ exhibiting no $f$-dispersion, as displayed in the upper inset. The magnitude is considerably high due to high conductivity and it decreases with increasing $f$ since by definition, $\epsilon^{\prime\prime} \propto 1/\omega$. In addition to $T_{FE}$, another well defined peak in $\epsilon^{\prime}(T)$ is observed at high temperature displaying $f$-dispersion, which will be discussed later in the context of Figs. \ref{Mem}(a) and \ref{Mem}(b).

$P(T)$, displayed in Fig. \ref{Dielec}(b), is measured at a poling field of $\pm$0.98 kV/cm. $P(T)$ develops non-zero value much above $T_N$ and gradually rises up to the lowest temperature, a behavior similar to that observed in AgCrS$_2$.\cite{Singh} Magnitude of $P(T)$ is significantly larger than AgCrS$_2$, despite of considerable conductivity ($\sim$ 10$^{-3}$ Scm$^{-1}$) of CuCrS$_2$.\cite{Singh} Neutron diffraction study, reported earlier, reveals a spiral spin order of the Cr$^{3+}$ ions in an AFM ground state with an incommensurate propagation vector for CuCrS$_2$.\cite{Str-ref, Rasch, Winten} The spiral spin arrangement breaks the inversion symmetry and causes ferroelectricity. Theoretical interpretation, based on the influence of spin-orbit interaction on the $d-p$ hybridization of ligand and transition metal orbitals, \cite{Jia} predicts ferroelectricity in the oxide counterpart, CuCrO$_2$.\cite{Seki} Requirement of the orbital hybridization has also been argued for interpreting ferroelectricity in AgCrS$_2$. We anticipate a similar mechanism of FE order in CuCrS$_2$. Nevertheless, the exact microscopic theory of ferroelectricity in the context of triangular lattice driven spin frustration is still under debate.

\par
The magnetodielectric percent (MD\%), defined as $[\epsilon^\prime(H) - \epsilon^\prime(0)]/\epsilon^\prime(0) \times 100$, is shown as a function of $T$ at $H$ = 10 (circle) and 70 (triangle) kOe in Fig. \ref{Dielec}(c), measured at $f = 900$ kHz. The data shows a substantial positive magnitude with two broad maxima, one below $T_N$ and another in the range, $T_N < T < T_{SR}$. It is customary to investigate the possibility of the origin of MD from magnetoresistance (MR) due to MW effect. MR\%, defined as $[\rho(H) - \rho(0)]/\rho(0) \times 100$, is plotted with $T$ in Fig. \ref{Dielec}(d) (right axis) in the same scale as MD\% for comparison. We observe a negative MR with a well defined peak below $T_N$ around which the low-$T$ peak in MD is  evident. The magnitude of MR is smaller compared to the significant value of MD and the result indicates a minor contribution to MD. We have also plotted magnetoloss (ML), defined as $[\tan\delta(H) - \tan\delta(0)]/\tan\delta(0) \times 100$, at $H$ = 10 (circle) and 70 (triangle) kOe in Fig. \ref{Dielec}(d) (left axis). The nature of ML\% is similar to MD\% while broad peaks appear at approximately the same temperatures as MD. In a $T$-dependence, the coincidence of the peak of $\tan\delta(T)$ and the inflection point of $\epsilon^\prime(T)$ indicates response of permanent electric dipoles.\cite{Kao,tand} This is manifested as coincidence of peaks in the difference plots, $\Delta\epsilon^\prime(T)$ and $\Delta\tan\delta(T)$. Hence, although MR shows a peak, the significant MD below $T_N$ thus  involves magnetic response of the permanent electric dipoles. Absence of any peak in MR above $T_N$, as evident in Fig. \ref{Dielec}(d), undoubtedly indicates that the high-$T$ MD effect does not appear due to MR, rather it involves ME coupling.   

\par
A strong $f$-dependence is observed at the high-$T$ peak ($T_m$) in $\epsilon^\prime(T)$, as evident in Fig. \ref{Dielec}(a). $T_m$ shifts toward high-$T$ with increasing $f$. Firstly, peak in the {\it real} permittivity indicates a kind of electric order. Secondly, the $f$-dependence of $T_m$ indicates distribution in the response of fluctuating entities $-$ short range nature of the order. Thirdly, broadness and large shift of the peaks are the indication of slow response and large inertia of the fluctuating entities $-$ possibly large volumes of the entities. The $f$-dependence of $\epsilon^\prime(T)$ can be satisfactorily analyzed either by Vogel Fulcher (VF) law or Dynamic scaling (DS) model, presented in Figs. \ref{Mem}(a) and \ref{Mem}(b), respectively. Satisfactory fit using VF law [Fig. \ref{Mem}(a)], defined by $f(T_m) = f_0 \exp[E/\{k_B(T_m-T_{g})\}]$, provides $T_{g} = 55$ K, $f_0 = 1.4 \times 10^7$ Hz, and $E = 34$ meV. On the other hand, satisfactory fit using the DS model, as shown in Fig. \ref{Mem}(b), defined by $f(T_m) = f_0 (T_m/T_{g} - 1)^{z\nu}$, gives $T_{g} = 90$ K, $f_0 = 6 \times 10^5$ Hz, and $z\nu = 2.64$. The value of $z\nu$ obtained from DS model is less than the values observed for spin-glasses\cite{manoj} while $f_0$ obtained from both the models are quite slower than the canonical spin glasses.\cite{Mydosh} However, agreement of the $f$-dependence of $T_m$ with both the laws signifies dipole glassiness, analogous to the scenario established in the spin sectors.\cite{Multi,klee,Karmakar-EPL}

\par
An essential requirement for designating glassy state is the exhibition of {\it memory effect} after isothermal ageing at a certain temperature below the glassy temperature ($T_g$). The memory effect is shown in Figs. \ref{Mem}(c)$-$(f) under different conditions. For all these cases the sample is annealed at a `wait' temperature ($T_w$) for 8 or 9 h ($t_w$) and then cooled to the lowest $T$ after which the data is recorded in the warming cycle. During annealing at a stable temperature for a long time, the ground state evolves, using thermal energy, through a set of gradually diminishing local energy minima separated by small energy barriers. When the sample is cooled further, the following energy states that evolve are characteristic or `branches' of the new ground state, rather than the states through which the system would have evolved without waiting. Hence, when the sample is reheated, the ground state retraces the new path until it passes through $T_w$, which is thus imprinted in the thermal dependence as a `memory' effect. The system is rejuvenated when the temperature reaches much above $T_w$. Figure  \ref{Mem}(c) demonstrates memory under zero magnetic field for $T_w =$ 22, 80, and 100 K and $t_{w} = 8$ h measured at $f = 4$ kHz. The data for $T_w = 22$ and 100 K do not show any memory while the data for $T_w = 80$ K shows a distinct sharp dip at 80 K demonstrating the memory effect. Absence of memory for waiting at 22 K signifies absence of glassy state below $T_N$ where long-range order prevails. However, memory effect for $T_w = 80$ K directs us to abandon the VF model, since it cannot explain the glassy behavior at 80 K. The DS model rather provides consistent information where we obtained $T_g$ = 90 K. In fact, we observe no memory in the curve for $T_w = 100$ K ($> T_g$) where glassy state no longer remains. 

\par
Figure \ref{Mem}(d) shows dependence of memory effect on $f$ for $T_w = 50$ K and $t_w = 8$ h. The data for  $f = 707$ Hz shows a 5 times stronger dip than that for $f = 3$ kHz indicating that the memory effect observed is stronger for low $f$ and is usually measured in low $f$. We, however, observe considerable memory even at high $f$. Figure \ref{Mem}(e) demonstrates memory for different $t_w$. It is clear that the dip at $T_w$ gradually increases for increasing $t_w$ (2, 4, and 8 h). The memory effect observed in various conditions strengthens our proposition of a dipole glass like state above $T_N$. Figure \ref{Mem}(f) shows $H$ dependence of the memory. The data in zero field shows the maximum dip which reduces considerably at 10 kOe and nearly vanishes at 100 kOe. The erasing of dielectric memory upon application of magnetic field signifies ME coupling, as also indicated by the significant MD effect. The results displaying the memory effect and magnetic field induced rejuvenation have huge potential both for technological applications and fundamental interest. The result implies either magnetic field induced transformation of the short range order to long range FE order or simply alignment of the net polarizations of short range clusters along magnetic field direction. This erases {\it disorder} in the dipole alignments, primarily required to build up glassy state. The former possibility can, however, be ruled out since the peak in $\epsilon^\prime(T)$ at $T_N$, signifying long range order of electric dipole moments, does not shift at all upon application of $H$, as evident in the inset of Fig. \ref{Dielec}(c). This is also apparent in Figs. \ref{Dielec}(c) and \ref{Dielec}(d) where sharp falls in MD\% and ML\% at $T_N$ show no shift at different $H$. Therefore, the second possibility seems plausible which gives rise to the magnetic field induced rejuvenation. Materials possessing such cluster dipolar regions exhibiting glassy dynamics are normally termed as relaxor ferroelectrics.

\par
The $\chi(T)$ and $C_P(T)$ behaviors indicate preformation of short range magnetic clusters in between  $T_{SR}$ (= 110 K) and $T_N$ (= 38 K). Each of these magnetic clusters bears a net electric polarization which is the resultant over the volume of the cluster. Each cluster thus acts as a large dipole or in a sense a `{\it super}' dipole. Because of random nucleation in zero magnetic field, the clusters magnetize in arbitrary directions and hence the polarizations of the clusters are oriented randomly. The randomness and frustration produce glassy behavior in the dielectric permittivity, akin to the cluster spin glasses.\cite{Mydosh} Notably, a low value of the attempt frequency ($f_0 \sim 10^5$ Hz), obtained from the analysis using DS law, signifies very slow response to the exciting signal resembling large entities rather than the independent electric dipoles. The material thus exhibits, what we call, a {\it super dipole glass} state which can be controlled magnetically. 

In addition to the technological importance, the observed memory and magnetic field driven rejuvenation are even appealing for  fundamental interest. The phenomenon should find importance, specially, in the issue of influence of topological frustration on the electronic degrees of freedom. It has been well accepted that a layered triangular lattice with AFM interactions typically gives rise to frustration in magnetism,\cite{Ramirez} which is indeed observed here. We have $T_N$ = 38 K, although a large $\left|\Theta_{CW}\right| (130$ K) is observed, demonstrating significant magnetic frustration. Emergence of frustration in the electric counterpart attributed to the geometrical constraint is very rare.\cite{Hemberger} This opens up a new dimension of ME coupling associated with glassy dynamics in the electronic degrees of freedom, complimentary to the well characterized field in the spin sector.\cite{Mydosh} In addition to the structural instability, the effect of covalency in chalcogenides seems to play a crucial role in the origin of {\it super dipole glass} state. For example, the oxide counterpart, CuCrO$_2$, also exhibits similar FE order at $T_N$,\cite{Seki} however it does not show any $f$ dispersion in the dielectric response.\cite{kiran} Thus, combined effects of structural instability and metal ligand hybridization must be taken into account for understanding these uncommon results. Proper theoretical interpretation along with more experimental findings with new paradigms is needed to be explored in the emerging field of glassiness correlated to multiferroicity.

\par
To our knowledge, evidence of dielectric memory and magnetic field induced erasing have not been reported earlier. Besides, the compound exhibits a rare consequence of relaxor multiferroicity, although study of which is challenged by its relatively high conductivity.

\vskip 0.1 in
\noindent
{\bf Acknowledgement}
S.G. acknowledges DST Nanoscience unit of IACS, Kolkata for MPMS facility. S.G. and S.C. also thank DST, India (Project no. IR/S2/PU-06/2006) for the high magnetic field facility. K.D. acknowledges CSIR, India for the SRF fellowship.

\end{document}